\documentclass[12pt]{article}
\pdfoutput=1
\usepackage{epsf,amsfonts,amssymb,epsfig,amsmath,mathtools,graphics,slashed}
\usepackage{hyperref,hep,graphicx,subfig}
\usepackage{rotating}

\newcommand{\nn}{\notag \\}

\makeatletter

\@addtoreset{equation}{section}
\makeatother

\begin{document}

\begin{titlepage}

\vfill

\begin{flushright}
Imperial/TP/2016/JG/02\\
DCPT-16/35
\end{flushright}

\vfill

\begin{center}
   \baselineskip=16pt
   {\Large\bf Holographic thermal DC response\\ in the hydrodynamic limit}
  \vskip 1.5cm
  \vskip 1.5cm
Elliot Banks$^1$, Aristomenis Donos$^2$, Jerome P. Gauntlett$^{1,3}$\\ Tom Griffin$^1$ and Luis Melgar$^1$\\
     \vskip .6cm
      \begin{small}
      \textit{$^1$Blackett Laboratory, 
  Imperial College,  London, SW7 2AZ, U.K.}
        \end{small}
         \vskip .6cm
      \begin{small}
      \textit{$^2$Centre for Particle Theory and Department of Mathematical Sciences\\Durham University, Durham, DH1 3LE, U.K.}
        \end{small}\\
             \vskip .6cm
               \begin{small}
      \textit{$^3$School of Physics, Korea Institute for Advanced Study, Seoul 130-722, Korea}
        \end{small}\\   
\end{center}

\vfill

\begin{center}
\textbf{Abstract}
\end{center}
\begin{quote}

We consider black hole solutions of Einstein gravity that describe deformations of CFTs at finite temperature 
in which spatial translations have been broken explicitly. We focus on deformations that are periodic in the non-compact
spatial directions, which effectively corresponds to considering the CFT on a spatial torus with a non-trivial metric.
We apply a DC thermal gradient and show that in a hydrodynamic limit
the linearised, local thermal currents can be determined by solving linearised, forced Navier-Stokes equations for an incompressible fluid on the torus. We also show how sub-leading corrections to the thermal current can be calculated 
as well as showing how the full stress tensor response that is generated by the DC source can be obtained. We also compare our results with the fluid-gravity approach. 
\end{quote}

\vfill

\end{titlepage}

\setcounter{equation}{0}
\section{Introduction}

In studying a strongly coupled system using holographic techniques an important and fundamental observable to study is the
response of the system to an applied DC thermal or electric source. In order to get finite results one needs, in general, a mechanism to dissipate momentum. Within holography the most natural way to achieve this is via the framework of holographic lattices \cite{Hartnoll:2012rj,Horowitz:2012ky,Donos:2012js,Chesler:2013qla,Donos:2013eha,Andrade:2013gsa}. 
These are stationary black hole geometries dual to CFTs in thermal equilibrium that have been deformed by marginal or relevant operators that explicitly
break the translation invariance of the CFT. Several different kinds of holographic lattices have now been studied and they give rise to a wide variety of interesting phenomena including Drude physics \cite{Hartnoll:2012rj,Horowitz:2012ky,Donos:2014yya}, novel metallic ground states \cite{Donos:2014uba,Gouteraux:2014hca}, metal-insulator transitions \cite{Donos:2012js,Donos:2013eha} and
anomalous temperature scaling of the Hall angle \cite{Blake:2014yla}.

It is now known that the DC thermoelectric conductivity matrix 
can be obtained for holographic lattices, universally, by solving a system of linearised Navier-Stokes equations for an auxiliary incompressible fluid on the black hole horizon. This was first shown in \cite{Donos:2015gia} and then extended in \cite{Banks:2015wha,Donos:2015bxe}. In a nutshell, this works as follows. By solving the fluid equations, which we also refer to as Stokes equations, one obtains
local thermal and electric currents on the black hole horizon, at the level of linear response, as functions of the applied DC sources. 
Furthermore, the total thermal and electric current fluxes at the horizon
are not renormalised in the radial direction and hence one also obtains the total thermal current fluxes of the dual field theory and thus the DC conductivity matrix. The results of \cite{Donos:2015gia,Banks:2015wha,Donos:2015bxe} have recently been used to obtain interesting bounds on DC conductivities \cite{Grozdanov:2015qia,Grozdanov:2015djs}.

We now elaborate a little further on several important aspects of the results of \cite{Donos:2015gia,Banks:2015wha,Donos:2015bxe}, focussing on static holographic lattices. Firstly, although not essential, it is helpful to introduce the DC sources via perturbations that are linear in time. This enables one to directly parametrise the thermal and electric DC sources using globally defined one-forms, 
rather than locally defined functions. This is helpful in constructing a globally defined bulk perturbation, 
and hence extracting the DC response. Such linear in time sources for a DC electric field and thermal gradient
were first discussed in \cite{Donos:2014uba} and \cite{Donos:2014cya}, respectively.

The second aspect we want to highlight is the result of \cite{Donos:2015gia} that the total thermal and electric current fluxes at the horizon are equal to the current fluxes at infinity. 
This was first observed for the special case of electric currents in the early paper \cite{Iqbal:2008by}, extending \cite{Kovtun:2003wp},
but in a very simplified setting. In particular, the black holes of \cite{Iqbal:2008by} did not incorporate momentum dissipation and the finite electric DC conductivity arose because the background black holes were electrically neutral, giving rise to constant electric currents. In fact, the thermal DC conductivity, which was not discussed in \cite{Iqbal:2008by},
is infinite for these black holes. For general holographic lattices, both the electric and thermal currents have non-trivial spatial dependence and it is only the total current fluxes that are non-renormalised in moving from the horizon to the AdS boundary.
This general result, articulated in \cite{Donos:2015gia}, came in a series of stages, starting with studies of electric currents \cite{Donos:2014uba} and then for the more subtle case of thermal currents \cite{Donos:2014cya}, both for Q-lattice examples, followed by electric and thermal currents for inhomogeneous holographic lattices, with momentum dissipation in one spatial dimension \cite{Donos:2014yya}. 

The non-renormalisation of the total current fluxes is widely referred to as a
kind of membrane paradigm, following \cite{Kovtun:2003wp,Iqbal:2008by}. However, this is somewhat of a misnomer. A membrane paradigm, in this context, should determine the currents, or perhaps just the
current fluxes, on the black hole horizon
and hence the conductivity of the black hole horizon. However,
for a general holographic lattice, {\it a priori}, it could have been the case that in order to obtain
the current fluxes on the horizon, one would need to solve the full bulk equations of motion for the perturbation. If this was the case then one would not have been able to obtain the current fluxes of the dual field theory just from the horizon data and the notion of any kind of local membrane would be 
irrelevant. Happily this turns out not to be the case.
 
Indeed the third aspect we want to highlight, and logically distinct from the previous one, is that one can determine the currents at the horizon by solving a closed set of fluid equations, the Stokes equations, on the horizon \cite{Donos:2015gia}. 
The Stokes equations, which only involve a subset of the perturbation, can be obtained by considering a radial Hamiltonian decomposition of the bulk equations of motion and then imposing the constraint equations on a surface of constant radial coordinate, in the limit as one approaches the horizon. The fact that one has a closed set of equations on the horizon and that they arise from a set of constraint equations makes the result of \cite{Donos:2015gia} a version of the membrane paradigm, if one finds that terminology helpful. 
We also emphasise that this membrane paradigm is universal and makes no assumption about taking any hydrodynamic limit.
It is a striking fact that solving a hydrostatic problem gives rise to exact statements about correlation functions in
relativistic CFTs with momentum dissipation, without taking any hydrodynamic limit.

The two key stepping stones that led to the general result about Stokes equations were the results obtained for
Q-lattice examples \cite{Donos:2014uba,Donos:2014cya}  followed by the results for the more involved example of an inhomogeneous holographic
lattice with momentum dissipation in just one spatial direction \cite{Donos:2014yya}. In hindsight, the results of \cite{Donos:2014uba,Donos:2014cya,Donos:2014yya}, which were originally obtained by brute force, were possible
because they are two cases in which the Stokes equations can be solved analytically.
The Stokes equations also emphasise the fundamental role that is played by thermal currents in studying DC response. Indeed, even if one wants to obtain just a purely electric DC response in the field theory, in general one still must solve for both the thermal and electric currents at the horizon, in order to extract this information.  

For the special classes of one-dimensional holographic
lattices (i.e. there is only momentum dissipation in one spatial direction) and also for Q-lattices \cite{Donos:2013eha},
the DC conductivity can be explicitly solved in terms of the horizon data \cite{Donos:2015gia,Banks:2015wha,Donos:2015bxe}. For more general classes of lattices, there are two limiting situations where we can make 
some universal statements.
The first limiting situation is associated with what have been called ``perturbative lattices" and was discussed in 
\cite{Donos:2015gia,Banks:2015wha}. The second, and in general distinct, limiting situation
occurs in a long-wavelength, hydrodynamic limit and will be the main focus of this paper.

We first briefly discuss the perturbative lattices, which are associated with weak momentum dissipation, in order to contrast with the hydrodynamic limit.
By definition, perturbative lattices can be constructed perturbatively about a translationally invariant 
black hole solution using a small amplitude deformation. For example, imagine starting with
the AdS-Schwarzschild black brane solution or, if one wants to be at finite charge density, 
the AdS-Reissner-Nordstr\"om solution.
These solutions are dual to translationally invariant CFTs with no momentum dissipation and hence have infinite DC thermal conductivity. We then deform the dual field theory by marginal or relevant operators that depend on the spatial coordinates of the CFT.
It is assumed that the strength or amplitude of the deformation is fixed by a small dimensionless UV parameter $\lambda$. On the other hand we make no restriction on how the UV deformation depends on the spatial coordinates, and so we allow deformations with arbitrary wave numbers $k$.
Generically, for small $\lambda$, the UV deformation will not change the IR of the black hole geometry. As such the horizon geometry can be expanded perturbatively in $\lambda$ about the flat horizon. 

As shown in \cite{Donos:2015gia,Banks:2015wha}
one can then solve the Stokes equations on the horizon perturbatively in $\lambda$ to obtain expressions for
the currents and hence the DC conductivity. At leading order in the perturbative expansion
the currents at the horizon are of order $\lambda^{-2}$, which shows the DC response is parametrically large as expected for weak momentum dissipation. The leading order currents at the horizon are constant and by considering the full bulk perturbation one can further show that at leading order in $\lambda$ the local currents of the dual CFT are also constant. 

The final expressions for the currents depend on the corrections to the horizon data at order $\lambda^1$ .
While these are not explicitly known in terms of the UV data, one can still obtain some additional general results for the DC conductivity. For example, for the case of perturbative lattices which have non-vanishing charge density at leading order\footnote{If the charge density vanishes at leading order then one should expand about the AdS-Schwarzschild solution. In this case
the conductivity $\sigma$ will be of order $\lambda^0$.} one should expand about the AdS-Reissner-Nordstr\"om (AdS-RN) solution. The conductivities are then all proportional to the same matrix, of order $\lambda^{-2}$, and as shown in \cite{Donos:2015gia,Banks:2015wha}, extending \cite{Donos:2014cya}, this leads to a generalised Wiedemann-Franz law\footnote{This result was also be obtained via memory matrix techniques in
\cite{Mahajan:2013cja}, building on \cite{Hartnoll:2012rj}.}, $\bar\kappa/(\sigma T)=s^2/\rho^2$, as well as a simple expression for the
Seeback coefficient (thermopower) equal to $\alpha/\sigma=s/\rho$.

We now turn to the hydrodynamic limit of holographic lattices. We will study this in the context of holographic lattices in Einstein gravity without matter fields in arbitrary spacetime dimensions $D\ge4$. As such our analysis is applicable to all CFTs in $D-1$ spacetime dimensions which have a classical gravity dual. The deformed CFTs can either be viewed as arising from deformations associated with the stress-energy tensor operator or, equivalently, by placing the CFT on a curved geometry. 

If we let $k$ be the largest wave number associated with the spatial deformations in the holographic lattice then
we consider the hydrodynamic limit\footnote{In the bulk we will take the limit $k/r_H<<1$, where $r_H$ is the radial location of the black hole. We can thus also consider the hydrodynamic limit as $k/s^{1/(D-2)}<<1$ where $s$ is the entropy density.} $\epsilon=k/T <<1$. In general this is distinct from the 
perturbative lattice. This can be seen from the following very explicit example. Consider a CFT at finite $T$ on a conformally flat spatial metric $h_{ij}=\Phi(x)\delta_{ij}$ with $\Phi(x)=\lambda \cos kx$. The perturbative lattice takes $\lambda<<1$
whereas the hydrodynamic limit takes $k/T<<1$. Note, however, that in both limits $\lambda\to 0$ and $\epsilon\to0$ we
end up with the AdS-Schwarzschild black brane. 

A key simplification which we show occurs in the hydrodynamic limit is that the black hole horizon geometry can be simply expressed explicitly in terms of the UV data. 
Thus, to obtain the DC linear response we solve the Stokes equations on a geometry that is explicitly known. Furthermore, we will show that at leading in order in the 
perturbative expansion in $\epsilon$, the local thermal currents at the horizon that are not renormalised in moving to the boundary. In other words, by solving the Stokes equations on the horizon one obtains the leading order local currents that are produced by the DC source,
at the level of linear response. The fact that there can be non-trivial local currents in the hydrodynamic limit
at leading order should be contrasted with the constant currents in the perturbative lattices. Conversely, similar
to the perturbative lattices, the thermal conductivity diverges like $\epsilon^{-2}$ at fixed $T$ as $\epsilon\to 0$
and we will see that the limit is again associated with weak momentum dissipation.

It is also interesting to develop the perturbative hydrodynamic expansion in more detail and we do this 
to determine the 
corrections of the thermal currents to the DC linear response that
are sub-leading in $\epsilon$. In addition
we will also show that it is possible to solve for the full radial dependence of the leading order bulk, linearised perturbations after applying a DC thermal source. This allows us to extract expressions not only for the local
thermal currents but also for the full linearised stress tensor of the boundary theory as a function of the applied DC thermal source. 
As one might expect, the stress tensor takes the form of constitutive relations for a forced fluid and after imposing the Ward identities one obtains, at leading order in the $\epsilon$ perturbative expansion, the non-relativistic Stokes equations.
The calculation also indicates how the sub-leading corrections to the stress tensor, perturbative in $\epsilon$, can be obtained in terms
of solutions to the Stokes equations on the horizon geometry, which is also expanded in $\epsilon$. As we shall see, our results also reveal a subtle interplay between sub-leading corrections in the perturbative expansion, solutions of the Stokes equations and regularity of the bulk perturbation at the horizon.

It is natural to ask how our results relate to work on fluid-gravity \cite{Bhattacharyya:2008jc} which relates
perturbative solutions of gravity equations to hydrodynamic equations.
In particular, CFTs on curved manifolds
have been studied in the context of theories of pure gravity from a fluid-gravity perspective in \cite{Bhattacharyya:2008mz}.  
Using a Weyl covariant formalism a hydrodynamic expansion ansatz for the bulk metric was presented. The ansatz gives rise to boundary stress tensor which, when the Ward identities are imposed, leads to perturbative solutions of the bulk Einstein equations.
One might expect our perturbative expansion in $\epsilon$, driven by the linearised DC source is a special case of the expansion
in \cite{Bhattacharyya:2008jc}. However, we show that off-shell, i.e. without imposing the Ward identities, the expansions differ, but they do agree on-shell. The origin of this difference seems to be related to the fact that we use Schwarzschild type coordinates (as in the hydrodynamic expansion of \cite{Janik:2005zt} in a different context) and impose the Ward identities at the black hole horizon (which are the Navier-Stokes equations),
while \cite{Bhattacharyya:2008jc} uses Eddington-Finklestein type coordinates and imposes the Ward identities at
the AdS boundary.

Before concluding this introduction we note that there have been various other studies of hydrodynamics in the context of momentum dissipation in holography, 
including \cite{Davison:2013txa,Balasubramanian:2013yqa,Davison:2014lua,Davison:2015bea,Blake:2015epa,Blake:2015hxa,Burikham:2016roo}. We highlight that a fluid gravity expansion was developed in 
\cite{Blake:2015epa,Blake:2015hxa} for the special sub-class of holographic lattices with spatial translations broken by bulk massless axion fields that depend linearly on the spatial coordinates. 
We also note that general results on hydrodynamic transport for quantum critical points have been presented in, for example, \cite{Hartnoll:2007ih,Forcella:2014gca,Lucas:2015lna,Lucas:2015sya}.
In contrast to the approach of \cite{Lucas:2015lna} we will see that within holography in the hydrodynamic limit we need to solve covariantised Navier-Stokes equations with constant viscosity.

\section{Holographic lattices in the hydrodynamic limit}\label{hightsols}

In the bulk of this paper we consider theories of pure gravity in $D$ bulk spacetime dimensions 
with action given by
\begin{align}
S=\int d^Dx\sqrt{-g}\left[R+(D-1)(D-2)\right]\,.\label{bulkaction}
\end{align}
We have set $16\pi G=1$ for simplicity. We have also chosen the cosmological constant so that a unit radius
$AdS_D$ spacetime solves the equations of motion.
The holographic lattice solutions which we study are static black holes that lie within
the ansatz:
\begin{align}\label{ansatz}
ds^{2}=-U\,G\,dt^2+\frac{F}{U}\,dr^{2}+g_{ij}dx^i dx^j\,.
\end{align}
Here $G,F$ and the $d= (D-2)$-dimensional spatial metric, $g_{ij}$, 
can depend on both the radial coordinate $r$ and the spatial coordinates of
the dual field theory, $x^{i}$. The function $U$ is a function of $r$ only, and is also
included for convenience. Thus,
\begin{align}
G=G(r,x),\quad F=F(r,x),\quad g_{ij}=g_{ij}(r,x),\quad U=U(r)\,.
\end{align}

We assume that the solutions have a single black hole Killing horizon located at $r=r_H$. 
It will be helpful to introduce another radial coordinate $\rho=r/r_H$, so that the horizon is located at $\rho=1$, 
and choose
\begin{align}
U&=r_{H}^2\,\rho^{2}\,u(\rho)\,,\qquad u(\rho)=1-\rho^{1-D}\,.
\end{align}
The temperature of the black hole is given by 
$4\pi T={(D-1)\,r_{H}}$ and we note that regularity of the metric at the horizon imposes the condition $F|_{\rho=1}=G|_{\rho=1}$.

The $AdS$ boundary is located at $r\to \infty$ where we demand that
\begin{align}
G\to h_{tt}(x)\,,\qquad g_{ij}\to r_{H}^2\,\rho^{2}h_{ij}(x)\,,
\end{align}
corresponding to studying the dual CFT on the curved background with metric $ds^2=-h_{tt} dt^2+h_{ij}dx^i dx^j$. Equivalently,
we are considering the CFT with deformations of the stress tensor with sources parametrised by $h_{tt}$ and $h_{ij}$.
We will focus on cases in which the spatial sections, parametrised by the $x^i$ are non-compact with planar topology.
Furthermore, we assume the deformations are periodically modulated in the spatial directions with associated wave numbers 
in the $x^i$ direction to be an
integer multiple of a minimum $2\pi/L^i$. Effectively, this
means that we can take the $x^i$ to parametrise a  torus with $x^i\sim x^i+L^i$. 

To obtain these holographic lattice solutions one needs to solve the non-linear PDEs subject to these boundary conditions. 
Some explicit examples have appeared in \cite{Balasubramanian:2013yqa,Donos:2014gya} with the construction in \cite{Donos:2014gya} exploiting a residual symmetry leading to 
solving a system of ODEs.
In this paper we will be interested in studying the hydrodynamic limit of these solutions. More precisely if we suppose $k$ is the largest
wave number associated with the spatial deformations then we are interested in studying the limit $\epsilon\equiv k/T<<1$. Equivalently, for these black holes we can consider the limits $k/r_H<<1$ or $k/s^{1/(D-2)}$ where $s$ is the entropy density. In fact, when solving the bulk equations of motion the latter are more natural.

By directly solving Einstein equations we find that the leading order solution is given by
\begin{align}\label{eq:gen_back}
ds^2&=\frac{(1+\rho\partial_\rho\ln w)^2}{\rho^2\,u(w \rho)}d\rho^2+r_H^2\rho^2 \frac{w^2}{h_{tt}(x)}\left[-u(w \rho) {h_{tt}(x)dt^2+{h_{ij}(x)}}dx^idx^j\right],
\end{align}
with $w(\rho, x)$ an arbitrary function satisfying
\begin{align}\label{wcond}
w(\rho, x)& \rightarrow 1,\quad \rho\rightarrow 1\,,\nn
w(\rho, x)& \rightarrow [h_{tt}(x)]^{1/2},\quad \rho\rightarrow \infty\,.
\end{align}
This solution solves the Einstein equations in the limit of ignoring spatial derivatives.
It should be viewed as the leading term in an asymptotic expansion in $\epsilon$, where
have neglected corrections of order $\epsilon^2$ as well as corrections that are non-perturbative in $\epsilon$.
 
For the remainder of the paper we will focus on the case with $w=1$ and hence consider the leading order
solution given by  
 \begin{align}\label{eq:gen_back2}
ds^2&=\frac{d\rho^2}{\rho^2\,u(\rho)}+r_H^2\rho^2 \left[-u(\rho) dt^2+h_{ij}(x)dx^idx^j\right]\,.
\end{align}
This corresponds to setting $h_{tt}=1$ and hence studying the dual CFT on the metric $ds^2=-dt^2+h_{ij}(x)dx^i dx^j$. In fact this is almost without loss of any generality. Indeed we can incorporate a non-vanishing $h_{tt}$ by simply performing a Weyl transformation of this boundary metric. Since we are considering CFTs the physics will be the same with the exception that in the case of odd $D$, {\it i.e.} when the CFT is in even spacetime dimensions, for certain metrics we will need to take into account the conformal anomaly, which will, in any case,
be a sub-leading effect in $\epsilon$.

Notice that the leading order solution \eqref{eq:gen_back2} is just the standard $AdS_D$ black brane solution but with the flat metric on
spatial sections, $\delta_{ij}$, replaced with the metric $h_{ij}$, which parametrises the UV deformations of the dual CFT. 
A corollary, which will be important in the following, is that the metric on the black hole horizon is given by $r_H^2h_{ij}(x^i)dx^idx^j$,
at leading order in the $\epsilon$ expansion. This accords with the intuition that at high temperatures the black hole horizon is approaching the AdS boundary.

It is straightforward to show that \eqref{eq:gen_back2} solves the Einstein equations, to leading order, 
after using radial coordinate $\rho$ and taking the limit $k/r_H\to 0$.
A simple way to derive \eqref{eq:gen_back} from \eqref{eq:gen_back2} is to simply rescale
the radial coordinate, $\rho\to w \rho$. If $w$ was a constant this would give a constant rescaling of the boundary metric. Taking $w$ to be a function of $(\rho,x)$ satisfying \eqref{wcond} leads to a Weyl transformation of the boundary metric at leading order in the
hydrodynamic limit. In particular this transformation introduces a
$d\rho dx^i$ cross term of order $\epsilon$, but after suitably shifting the $x^i$ coordinates one obtains the solution 
\eqref{eq:gen_back} dropping just $\epsilon^2$ corrections (of the corrections that are polynomial in $\epsilon$).
In general, from \eqref{eq:gen_back} we see that the metric on the black hole horizon 
in the hydrodynamic limit is given by
\begin{align}\label{bhhorgen}
ds^2_H=r_H^2\frac{h_{ij}}{h_{tt}}dx^i dx^j\,.
\end{align}
In particular, we emphasise that it is invariant under Weyl transformations of the boundary metric $h_{\mu\nu}\to e^{\gamma}h_{\mu\nu}$.

\section{Thermal currents from a DC source}\label{review}

The linear response that arises from the application of a DC thermal gradient can be calculated within holography by
studying a suitable linearised perturbation of the gravitational background. We first review some of the key results of \cite{Donos:2015gia,Banks:2015wha,Donos:2015bxe}, applicable to
all holographic lattices, before turning to the long wavelength limit. In particular, we show how the local thermal currents can be
obtained at leading order in the hydrodynamic limit.

\subsection{Review of general DC response via Stokes equations}
As first discussed in \cite{Donos:2014cya} it is convenient
to introduce the DC source by considering the following linearised perturbation about the general background holographic lattice geometry given in \eqref{ansatz} via:
\begin{align}\label{pertan}
\delta ds^2=-2UGt\zeta_i dt dx^i+\delta g_{\mu\nu} dx^\mu dx^\nu\,,
\end{align}
where $x^\mu=(t,r,x^i)$. The key piece here is the linear in time source that is parametrised by the one-form
$\zeta\equiv \zeta_i(x) dx^i$, which just depends on the spatial coordinates\footnote{Note that if we write $(x)$ the argument will always refer to the spatial field theory coordinates.} and is closed $d\zeta=0$. It is important to emphasise that $\zeta$ is a globally defined 
one-form on the boundary spacetime and can easily be shown to parametrise a thermal gradient. Indeed
if we write $\zeta=d\phi(x)$, for some locally defined function $\phi(x)$, after making the coordinate transformation
$t\to t (1-\phi)$ the linear perturbation appears in the perturbed metric as
\begin{align}\label{otherway}
ds^2+\delta ds^2=-(1-\phi)^2UGdt^2+\frac{F}{U}\,dr^{2}+g_{ij}dx^i dx^j+\delta g_{\mu\nu} dx^\mu dx^\nu\,,
\end{align}
and locally, we identify $\phi(x)=-\ln T(x)$ so that $\zeta=-T^{-1}dT$. One might consider, for example, $\phi(x)=\bar\zeta_i x^i$, with $\bar\zeta_i$ constant, and then we have $T^{-1}\partial_i T=-\bar\zeta_i$.

The remaining part of the perturbation, $\delta g_{\mu\nu}(r,x)$, is whatever is
required in order to get a consistent set of equations. 
The perturbation components $\delta g_{\mu\nu}$ 
are taken to be globally defined functions of the spatial coordinates\footnote{Note that in this paper we are considering solutions in which there is a single black hole horizon and in which the spatial coordinates continue from the boundary to the horizon. More general solutions are discussed in \cite{Donos:2015gia,Banks:2015wha,Donos:2015bxe}.}. They are chosen to suitably fall off fast enough at the AdS boundary to ensure that the only source being applied is parametrised by $\zeta$. The boundary conditions of
$\delta g_{\mu\nu}$ at the black hole horizon are chosen so that the full perturbation given in \eqref{pertan} is regular.
It is worth commenting that a helpful aspect of working with the globally defined one-form and time coordinate as in \eqref{pertan} is that it makes it clear that $\delta g_{\mu\nu}(r,x)$ should be globally defined functions. Indeed, {\it a priori}, it is not clear what conditions one should impose on the spatial dependence of the perturbations $\delta g_{\mu\nu}$ in \eqref{otherway} given the presence of the locally defined function $\phi$.

Following \cite{Donos:2015gia,Banks:2015wha} we introduce the {\it bulk} thermal current density, $Q^i(r,x)$, depending on both the radial direction and the spatial directions of the field theory, defined by
\begin{align}\label{defqueue}
Q^i(r,x)\equiv\frac{G^{3/2}U^2}{F^{1/2}}\sqrt{g_d}g^{ij}\left(\partial_r\left(\frac{\delta g_{jt}}{GU}\right)-\partial_j\left(\frac{\delta g_{rt}}{GU}\right)
\right)\,,
\end{align}
where $\sqrt{g_d}$ refers to the volume element of the $d$-dimensional spatial metric $g_{ij}(r,x)$ in \eqref{ansatz}.
By evaluating this at the $AdS$ boundary we obtain
\begin{align}
Q_{QFT}^i(x)&\equiv \lim_{r\to\infty} Q^i(r,x)\,,
\end{align}
where $Q_{QFT}^i(x)$ is the {\it local} thermal current density of the dual quantum field theory given by the stress-tensor, $Q_{QFT}^i(x)= -\sqrt{h}T^i{}_t(x)$, which is induced by the DC thermal gradient. We also define the local currents at the black hole
horizon via 
\begin{align}
Q_{BH}^i(x)&\equiv \lim_{r\to r_H} Q^i(r,x)\,.
\end{align}
It was shown in \cite{Donos:2015gia,Banks:2015wha} that the bulk thermal currents satisfy a differential equation which can be integrated in the radial direction leading to the following important relation:
 \begin{align}\label{quvhor}
Q^i_{QFT}=Q^i_{BH}-
\partial_j\int_{r_H}^\infty dr\left( 
(GF)^{1/2} \sqrt{g_d}g^{ik}_d g^{jl}_d
\Bigg((UG)\partial_k\left(\frac{\delta g_{lt}}{GU}\right)-k\leftrightarrow l\right)\Bigg)\,.
\end{align}
Notice that the second term on the right hand-side of this expression is a magnetisation current and
is trivially conserved.

Next, in a radial Hamiltonian formulation of the equations of motion, by evaluating the constraints at
the black hole horizon, one can show that a {\it subset} of the perturbation is governed by a system of forced linearised Navier-Stokes equations, also called Stokes equations,  
on the horizon:
\begin{align}\label{stokesgeneral}
-{2}\,{\nabla}^{Hi}{\nabla^H}_{\left(i\right.}v_{\left.j\right)}=4\pi T\zeta_{j}-\partial_{j}p\,,\qquad \nabla^H_i v^i=0\,,
\end{align}
where
\begin{align}\label{veepee}
v_i=-\delta g_{it}|_H,\qquad p=-\big(\delta g_{rt}\frac{4\pi T}{G}+\delta g_{it}g^{ij}\nabla_j\ln G\big)|_H\,,
\end{align}
and here $\nabla^H$ is the Levi-Civita connection associated with spatial metric on the horizon $g_{ij}|_H$.
Furthermore, the local thermal currents on the horizon are given by
\begin{align}\label{qexp}
Q^i_{BH}=4\pi T\sqrt{g_d}|_H g^{ij}_Hv_j\,.
\end{align}

It should be emphasised that, in general, the fluid on the horizon is an auxiliary fluid and only indirectly related to
observables in the boundary CFT. If we solve the Stokes equations on the horizon geometry, we obtain the local thermal
currents $Q^i_{BH}(x)$ from \eqref{qexp}. We can then obtain the physical local thermal currents $Q^i_{QFT}(x)$ from \eqref{quvhor},
but only if we have solved the full radial dependence of the perturbation in the bulk. 

However, still quite generally, if we have solved the Stokes equations we can easily obtain the total current flux $\bar Q^i_{QFT}$ and hence the DC thermal conductivity matrix
defined via $\bar Q^i_{QFT}=T\kappa^{ij}\bar \zeta_j$. To see this we define the 
total current flux, or equivalently, the zero mode of the current, via
\begin{align}
\bar Q^i_{QFT}\equiv\int Q^i_{QFT}\,,
\end{align}
where\footnote{This way of defining the zero modes is widely used. One could also define them by
averaging with $\int \Pi_i dx^i\sqrt{h}$.} $\int\equiv (\Pi_i L_i)^{-1}\int \Pi_i dx^i$ refers to an average integral over a period in the spatial directions. 
Defining $\bar Q^i_{BH}$ in a similar way, we can immediately deduce from \eqref{quvhor} that 
$\bar Q^i_{QFT}=\bar Q^i_{BH}$.

\subsection{DC thermal current response for $k/T<<1$}\label{crtresults}
We now consider the universal results for holographic lattices that we just summarised in the context of the 
hydrodynamic limit. We will consider boundary metric given by $ds^2=-dt^2+h_{ij}(x)dx^i dx^j$.
There are two key simplifications.

The first is that when $\epsilon=k/T<<1$ the black hole horizon metric  
can be expressed explicitly in terms of the physical UV deformation of the dual CFT. 
Indeed from the leading order form of the solution given in \eqref{eq:gen_back2} we deduce that 
the black hole horizon metric is given by 
$r_H^2h_{ij}dx^idx^j$, where $h_{ij}(x)$ is the UV deformation of the dual CFT.
Thus, we know the explicit geometry for which we need to solve the Stokes equations \eqref{stokesgeneral} in order
to obtain the local currents on the black hole horizon $Q^i_{BH}(x)$.

The second simplification is that the second term on the right hand side of \eqref{quvhor} will be suppressed by
order $\epsilon^2$ and hence, at leading order, 
we deduce that the {\it local} currents at the horizon are the same as those in the dual field theory:
\begin{align}\label{localcurs}
Q^i_{QFT}(x)=Q^i_{BH}(x)+\mathcal{O}(\epsilon^2)\,.
\end{align}

Putting these two facts together we draw the following important conclusion.
Consider a holographic lattice describing a CFT on a curved manifold with arbitrary spatial metric $h_{ij}(x)$.
After applying a DC thermal source, parametrised by the closed one-form $\zeta$, 
at leading order in $\epsilon$ we can obtain the local thermal currents $Q^i_{QFT}(x)$ of the dual field theory
by solving the Stokes equations \eqref{stokesgeneral} using the metric $r_H^2h_{ij}dx^idx^j$ where $r_H=4\pi T/(D-1)$.
In the next section we will show how to obtain the full stress tensor of the fluid can be obtained at leading
order in $\epsilon$.

It is worth highlighting that in the hydrodynamic limit we are therefore solving the covariantised Navier-Stokes
equations with metric $r_H^2h_{ij}dx^idx^j$ and with constant viscosity given by $\eta=r_H^{D-2}$. This should
be contrasted with the speculation in section 4 of \cite{Lucas:2015lna} that within holography one should consider 
spatially dependent viscosity and a conformally rescaled metric on the horizon. In fact it is also worth recalling here
a point made earlier that a Weyl transformation of the boundary metric $h_{\mu\nu}$ leads to the same black hole horizon metric, as we see from \eqref{bhhorgen}.

We conclude this section by deriving the scaling behaviour of the thermal conductivity in the hydrodynamic limit.
We can remove all dimensionful quantities from the Stokes equations by scaling $\hat x=k x$, $\hat p=kp$, $\hat\zeta=T\zeta$
and $\hat v^i=k^2 h^{ij}v_j$. Indeed the Stokes equations \eqref{stokesgeneral}
can then be written in terms of the hatted variables as well as the UV
metric deformation $h_{ij}$. We thus deduce that $\hat v^i$ is related to $\hat \zeta_i$ via a dimensionless matrix. From
\eqref{qexp}, and recalling that the metric on the horizon is $r_H^2h_{ij}dx^idx^j$, we deduce that the heat current scales as 
scaling $Q^i_{QFT}\propto T^{D-1}/k^2\sqrt{h}\hat v^i$ and thus we deduce that in the hydrodynamic limit we 
have\footnote{In \cite{Banks:2015wha} it was shown how one can solve the Stokes equations explicitly and hence obtain the thermal conductivity
$\kappa^{ij}$,
for all temperatures, if the metric deformation $h_{ij}$  only depends on one of the spatial coordinates. From the formulae in \cite{Banks:2015wha},
one can check that the high temperature limit of the result for $\kappa$ is indeed of the form \eqref{hightscale}.}.
\begin{align}\label{hightscale}
{\kappa}\propto \frac{s}{T\epsilon^{2}}\,,
\end{align}
where $s\propto T^{D-2}$ is the entropy density. Clearly at fixed $T$, this is parametrically large 
as $\epsilon\to 0$ as one expects for weak momentum dissipation.

\section{The full perturbation for the thermal DC source}\label{long}

In this section we examine the full perturbation that is induced by the thermal DC source
for a holographic lattice in the hydrodynamic limit $\epsilon=k/T<<1$.
This will allow us to obtain not only the local heat currents that are produced but also the full local stress tensor of
the dual field theory. In addition, this analysis will display in more detail the structure of the perturbative expansion
in $\epsilon$.

The strategy is to first solve the full radial dependence of the linearised perturbation about the background geometry that is associated with a DC thermal source parametrised by the closed one-form $\zeta$. 
At leading order in $\epsilon$ we obtain a local stress tensor that takes the form of constitutive relations 
for a fluid. Imposing the boundary Ward
identities, or equivalently the constraint equations with respect to a radial Hamiltonian
decomposition, then implies that the stress tensor satisfies the Stokes equations given in \eqref{stokesgeneral}. The details of how the perturbation scales with $\epsilon$ is somewhat subtle as we shall see. 

\subsection{The thermal gradient source on the boundary}\label{tgsotb}
We begin by considering a static boundary metric of the form $ds^2=-dt^2+h_{ij}(x) dx^i dx^j$, where the spatial metric $h_{ij}(x)$, which depends periodically on the spatial coordinates $x^i$, parametrises the holographic lattice. As above, the
thermal gradient source can be introduced by writing $\zeta=d\phi(x)$, for some locally defined function $\phi(x)$, and then considering the perturbed
boundary metric:
\begin{align}
ds^2=-(1-\phi)^2dt^2+h_{ij}dx^i dx^j\,.
\end{align}
We identify $\phi(x)=-\ln T(x)$ so that $\zeta=-T^{-1}dT$. 
If we make the coordinate transformation $t\to t+t\phi(x)$ we obtain
\begin{align}
ds^2=-dt^2+h_{ij}dx^i dx^j-2 t\zeta_i dx^i dt\,,
\end{align}
and we see the `linear in time source'.
Note that the perturbed metric is now expressed in terms of the globally defined one-form $\zeta$ and not the locally defined
function, $\phi(x)$.

In carrying out the calculations presented below, we found it convenient to work with a Weyl transformed
version of this metric, with Weyl factor given by $(1+p/(4\pi T))^2$ where $p(x)$ is a globally defined (periodic) function. 
At linearised order, we therefore consider
\begin{align}\label{anotbf}
ds^2=(1+\frac{2p}{4\pi T})\left[-dt^2+h_{ij}dx^i dx^j\right]-2 t\zeta_i dx^i dt\,.
\end{align}
If we now employ the additional coordinate transformation $t\to t(1-p/(4\pi T))$ we obtain
\begin{align}\label{pertscez}
ds^2=-dt^2+(1+\frac{2p}{4\pi T})h_{ij}dx^i dx^j-2 t[\zeta_i-(4\pi T)^{-1}\partial_i p] dx^i dt\,.
\end{align}
An appealing feature of using these coordinates is that the thermal gradient source is now appearing in the
combination $\zeta_i-(4\pi T)^{-1}\partial_i p$, as in the Stokes equations \eqref{stokesgeneral}. 

We will employ one further coordinate change by taking  $x^i\to x^i-th^{ij}\xi_j$, where $\xi_j=\xi_j(x)$, to finally write the perturbed metric in the form
\begin{align}\label{pertsce}
ds^2&=-dt^2+h_{ij}dx^i dx^j -2 t\left(\zeta_i-(4\pi T)^{-1}\partial_i p\right) dx^i dt\nn
&-2\left(\xi_{j}\,dt+t\,{\nabla}_{(i}\xi_{j)}\,dx^{i}\right)\,dx^{j}+\frac{2p}{4\pi T}h_{ij}dx^i dx^j\,.
\end{align}
Despite obscuring the fact that the perturbation corresponds to a simple thermal gradient, we found this way of
writing things helpful in extending the perturbation into the bulk as we discuss next.

\subsection{The linearised perturbation: solving the radial equations}
Writing the source on the boundary in the form given in \eqref{pertsce} is rather non-intuitive. However, it
has two virtues when we extend it into the bulk. The first is that it allows us to work in a radial gauge with $\delta g_{rt}=\delta g_{ri}=0$.
The second is that it helps to obtain a perturbation that is regular at the black hole horizon.

Recall from section \ref{hightsols}
that the leading order bulk background solution, at leading order in $\epsilon$, with boundary metric $ds^2=-dt^2+h_{ij}dx^i dx^j$,
is given by
 \begin{align}\label{eq:gen_back22}
ds^2&=\frac{d\rho^2}{\rho^2\,u(\rho)}+r_H^2\rho^2 \left[-u(\rho) dt^2+h_{ij}(x^i)dx^idx^j\right]\,.
\end{align}
Furthermore, this has corrections at order $\epsilon^2$.
We begin by considering the following bulk linearised perturbation that is induced by the applied source $\zeta_i(x) dx^i$:
\begin{align}
\delta ds^{2}&=-2\,r_{H}^{2}\rho^{2}\,u\,t\,[\zeta_{i}-(4\pi T)^{-1}\partial_{i}p ]\,dx^{i}\,dt\label{eq:pert_l1}\\
&-2r_H^2\rho^{2}\,(\xi_{j}\,dt+t\,{\nabla}_{(i}\xi_{j)}\,dx^{i})\,dx^{j}\label{eq:pert_l2}\\
&+r_H^2\rho^22\frac{p}{4\pi T}h_{ij}dx^i dx^j\label{eq:pert_l25}\\
&-2(\rho^{3-D}V_{j}+\dots)\,dx^{j}\,dt\label{eq:pert_l3}\\
&+\rho^{2}\,\frac{\ln u}{4\pi T}\,s_{ij}\,dx^{i}\,dx^{j}\label{eq:pert_l4}\,.
\end{align}
Here $p$, $\xi_{i}$, $V_{i}$, and $s_{ij}$ are all periodic functions of the boundary spatial coordinates $x^{i}$.
In this section we will raise and lower indices using the UV boundary metric $h_{ij}$ and 
$\nabla$ is the associated Levi-Civita connection. We also assume that $h^{ij}s_{ij}=0$.

We claim that each of the five lines separately solves the radial equations of motion
arising in the Einstein equations at leading order in $\epsilon$, with corrections to each line of order $\epsilon^2$.
However, an important subtlety, which we will carefully discuss in more detail, is that to ensure the
perturbation is well defined at the black hole horizon when we consider these perturbations altogether, we will have to determine 
the order in the $\epsilon$ expansion at which the various terms in 
\eqref{eq:pert_l1}-\eqref{eq:pert_l4} first appear.

We now discuss the various terms in more detail. The first three lines \eqref{eq:pert_l1}-\eqref{eq:pert_l25} are the radial extensions of the boundary source terms that we discussed in \eqref{pertsce}. The first line \eqref{eq:pert_l1} is almost the standard linear in time perturbation of the metric that we saw in \eqref{pertan}.
By shifting the time coordinate 
one can then easily show that this solves the Einstein equations, with corrections of
order $\epsilon^2$. Notice that $\zeta_{i}-(4\pi T)^{-1}\partial_{i}p$ appears rather
than just $\zeta_{i}$ as in \eqref{pertan}. This change, which is associated with our source terms given in \eqref{pertsce}, allows us to work in a radial gauge.

The second line \eqref{eq:pert_l2} can locally be generated from the background solution \eqref{eq:gen_back22} via the coordinate transformation
\begin{align}
x^{i}&\rightarrow x^{i}-\,t\,h^{ij}\xi_{j}\,,
\end{align}
and it therefore satisfies, trivially, the equations of motion, with corrections of order $\epsilon^2$.
The third line \eqref{eq:pert_l25} can also be generated from the background solution \eqref{eq:gen_back22}. Indeed the spatial metric
in \eqref{eq:gen_back22} was arbitrary and so the perturbation \eqref{eq:pert_l25} is simply obtained by taking $h_{ij}\to (1+2\frac{p}{4\pi T})h_{ij}$.

Having explained how the source terms given in \eqref{pertsce} extend into the bulk, we now discuss the last two lines
\eqref{eq:pert_l3}, \eqref{eq:pert_l4}, which are needed in order to obtain a consistent perturbed metric. Neither of these lines are 
associated with source terms in the boundary theory; instead they are associated with the response to the DC thermal gradient. Both \eqref{eq:pert_l3} and \eqref{eq:pert_l4} are solutions of Einstein's equations at leading order in $\epsilon$. We will discuss
\eqref{eq:pert_l3} in detail in appendix \ref{log} since, as we discuss below, sub-leading corrections to this perturbation will play an important role when we analyse regularity of the perturbation at the black hole event horizon. 
We have explicitly added the dots in \eqref{eq:pert_l3} to emphasise this point.

For the line \eqref{eq:pert_l4} one requires
the condition $h^{ij}s_{ij}=0$ and, ignoring spatial derivatives of $s_{ij}$, one finds
the given radial dependence after solving the $ij$ component of the Einstein equations. Indeed if we consider the ansatz $\delta ds^2=A(\rho)s_{ij}(x)$ then we find that the function $A(\rho)$ satisfies
\begin{align}\label{raddep}
\partial_\rho[\rho^D u\partial_\rho(\rho^{-2} A)]=0\,,
\end{align}
and we have chosen the solution $A\propto \rho^2\ln u$ in order that it has no source at
infinity.
Corrections are again of order $\epsilon^2$.

Let us now examine the behaviour of the perturbed metric at at the AdS boundary and at the black hole horizon.
It is clear by construction that as we approach the AdS boundary at $\rho\to\infty$, the perturbation 
\eqref{eq:pert_l1}-\eqref{eq:pert_l4} gives rise to precisely the source terms in \eqref{pertsce}, associated with a thermal gradient
parametrised by the closed one-form $\zeta$.

To examine the issue of regularity on the horizon, located at $\rho=1$, we employ
the Eddington-Finklestein ingoing coordinate $v\sim r_H(t+\ln u/(4\pi T))$.
We find that we should impose 
\begin{align}\label{eq:hor_reg}
&\xi_{j}=-r_H^{-2}V_j,\quad s_{ij}=2\,{\nabla}_{(i}V_{j)}\,,
\end{align}
and, after recalling that we imposed $h_{ij}s_{ij}=0$, the latter implies the incompressibility condition $\nabla_iV^i=0$.
After making these identifications we see that the metric is almost regular at the horizon, 
but there is a remaining singular term of the form 
\begin{align}\label{eq:pert_l1again}
-2r_H\frac{\ln u}{4\pi T}(\zeta_{i}-(4\pi T)^{-1}\partial_{i}p )\,dx^{i}d\rho\,,
\end{align}
as $\rho\to 1$. 

Before returning to this crucial issue we see that writing the boundary metric in the form
\eqref{pertsce}, which arose from a boundary coordinate transformation, implies that additional modes had to be activated
in the bulk, namely \eqref{eq:pert_l3} and \eqref{eq:pert_l4} with \eqref{eq:hor_reg}, in order to get a regular perturbation
at the horizon. A closely related fact is that the boundary coordinate transformation leading to \eqref{pertsce} can be extended
smoothly into the bulk. Specifically, consider the bulk coordinate transformations
\begin{align}\label{bulkctfs}
t&\rightarrow t\,(1+(4\pi T)^{-1}p)+g(\rho,x^{i})\,,\nn
x^{i}&\rightarrow x^{i}+\,t\,h^{ij}\xi_{j}+g^{i}(\rho,x^{i})\,.
\end{align} 
acting on the perturbed metric \eqref{eq:gen_back22} combined with \eqref{eq:pert_l1}-\eqref{eq:pert_l4}
If $g$ and $g^i$ vanish fast enough as $\rho\to \infty$ the conformal boundary will be given in the form \eqref{anotbf}. Furthermore, 
the coordinate transformations are smooth as we approach the horizon, provided that we choose
$g(\rho,x^{i})\rightarrow p\,\frac{\ln u}{(4\pi T)^2}$ and
$g^{i}(\rho,x^{i})\rightarrow \,h^{ij}\xi_{j}\,\frac{\ln u}{4\pi T}$
as $\rho\to 1$.

We now return to the remaining divergence \eqref{eq:pert_l1again} at the horizon. To cancel this it is necessary to 
consider sub-leading terms in the expansion in $\epsilon$, both for the background and for the perturbation.
As we discuss in more detail in appendix \ref{log},
there is a delicate cancelation between the sub-leading terms in
the $\epsilon$ expansion of \eqref{eq:pert_l3} (denoted by dots) and \eqref{eq:pert_l1again}. 
We find that regularity of the perturbation at the horizon implies that the leading order pieces of the perturbation
have a dependence on $\epsilon$ given by
\begin{align}
V_i(x)&=\epsilon^{-2}v_i^{(0)}(x)\,,\nn
p(x)&=\epsilon^{-1}p^{(0)}(x)\,,
\end{align}
with $v_i^{(0)}$ and $p^{(0)}$ the same order
as $4\pi T\zeta_i$. We emphasise that this behaviour is fixed by regularity at the horizon.

Furthermore, the first line \eqref{eq:pert_l1}
and the fourth line \eqref{eq:pert_l3} of the perturbation read 
\begin{align}\label{twolines}
\delta ds^{2}=&-2\,r_{H}^{2}\rho^{2}\,u\,t\,(\zeta_{i}-(4\pi T)^{-1}\epsilon^{-1}\partial_{i}p^{(0)} )\,dx^{i}\,dt\nn
&-2\epsilon^{-2}\rho^{3-D}(v_i^{(0)}(x)\,+\epsilon^2 V^{(2)}_i(\rho,x))\,dx^{j}\,dt\,,
\end{align}
where an explicit expressions for $V_i^{(2)}(\rho,x)$is given in appendix \ref{log} (see \eqref{fexac2}). Note that we have only explicitly written the 
one sub-leading term, $V_i^{(2)}$, which is relevant to the discussion here. As we approach the horizon
the sub-leading term behaves as
\begin{align}\label{veetwo}
V^{(2)}_i(\rho,x) \to \frac{r_H^2}{(4\pi T)^2}u \log(\rho-1)\left( -\frac{2}{(\epsilon r_{H})^{2}}\,{\nabla}^{i}{\nabla}_{\left(i\right.}v^{(0)}_{\left.j\right)}   \right), \qquad \rho\to 1\,.
\end{align}
We now see that the singular term \eqref{eq:pert_l1again} in the combined perturbation, arising from the first line
in \eqref{twolines}, is cancelled by the sub-leading term in the second line of \eqref{twolines} providing that
$v^{(0)}_i$ satisfies the Stokes equations,
\begin{align}\label{stokeslbulk}
-\frac{2}{(\epsilon r_{H})^{2}}\,{\nabla}^{i}{\nabla}_{\left(i\right.}v^{(0)}_{\left.j\right)}=4\pi T\zeta_{j}-\epsilon^{-1}\partial_{j}p^{(0)}\,.
\end{align}
It is satisfying to see that these equations are none other than the leading order expansion in $\epsilon$ of \eqref{stokesgeneral}, where we should recall that the black hole horizon metric is given, at leading order in $\epsilon$, by $(r_H)^2h_{ij}dx^i dx^j$. 
We thus conclude that there is a subtle interplay between sub-leading terms in the expansion, regularity of the perturbation
at the horizon and the Stokes equations.

\subsection{The boundary stress tensor}\label{stress}
We now calculate the holographic stress tensor for the dual field theory at leading order in the $\epsilon$ expansion.
We also present the sub-leading contributions to the local heat current, based on the calculations carried out in appendix 
\ref{log}.

We will express the result for the stress tensor in the coordinates and Weyl frame for the boundary metric given by
\begin{align}\label{pbmet}
ds^2=-dt^2+h_{ij}dx^idx^j-2t\zeta_idx^i dt\,.
\end{align}
To achieve this we first carry out the coordinate transformations\footnote{Note that since we are only interested in the boundary expansion here we don't need to consider the $g$ and $g^i$ terms in \eqref{bulkctfs}.}
\begin{align}\label{ctsfs}
t&\rightarrow t\,(1+(4\pi T)^{-1}p)\,,\nn
x^{i}&\rightarrow x^{i}+\,t\,h^{ij}\xi_{j}\,.
\end{align}
This leads to a perturbed metric 
which asymptotes to the boundary metric in \eqref{anotbf} with an overall Weyl factor given by $(1+\frac{2p}{4\pi T})$.
We can move to Fefferman-Graham type coordinates, which are a slight modification of those for AdS-Schwarzschild, and also
eliminate the boundary Weyl factor, via
\begin{align}
\rho&=\frac{ (1+\frac{1}{4}[r_Hz(1+(4\pi T)^{-1}p)]^{D-1})^{\frac{2}{D-1}}}{r_H z(1+(4\pi T)^{-1}p)}\,,\nn
&\approx\frac{1}{r_Hz(1+(4\pi T)^{-1}p)}\left(1+\frac{1}{2(D-1)}[r_Hz(1+(4\pi T)^{-1}p)]^{D-1}\right)\,,\qquad \rho\to\infty\,.
\end{align}
In these coordinates we have $g_{zz}=1/z^2$ and the AdS boundary is now located at $z=0$. Note that to the order 
in $\epsilon$ we are working with, we can drop spatial derivatives of $p$. The boundary conformal metric is now given by
\eqref{pbmet} and by determining the terms with factors $z^{D-3}/(D-1)$ a consideration of \cite{deHaro:2000xn} then implies that the stress tensor has components,
\begin{align}\label{stresspert}
T_{tt}&=(D-2)r_H^{D-1}+(D-2)r_H^{D-2}{p}\,,\nn
T_{ij}&=r_H^{D-1}h_{ij}+r_H^{D-2}\left(-r_H^{-2}2\nabla_{(i}V_{j)}+ph_{ij}\right)\,,\nn
T_{it}&=r_H^{D-1}\left[(D-2)\zeta_i t-(D-1) r_H^{-2}V_i\right]\,.
\end{align}
Recall from our discussion above that $V_i=\epsilon^{-2}v_i^{(0)}(x)$,
$p=\epsilon^{-1}p^{(0)}(x)$ and we recall that
$r_H=4\pi T/(D-1)$. Clearly the perturbed stress tensor takes the form of constitutive relations for a fluid with source
parametrised by $\zeta_i$. We should also note that we have already imposed the incompressibility
condition $\nabla_i V^i=0$.

When the constraint equations, in a radial Hamiltonian decomposition
of Einstein equations,
are imposed at the AdS boundary they imply that the stress tensor is traceless, $T^\mu{}_\mu=0$, and also satisfies
the Ward identity $\nabla_\mu T^{\mu\nu}=0$, both with respect to the deformed boundary metric \eqref{pbmet}. The first of these is already satisfied due to the incompressibility condition
\begin{align}
\nabla_i V^i=0\,,
\end{align}
and the second gives rise to the linearised Navier-Stokes equations: 
\begin{align}\label{sttwo}
-\frac{2}{( r_{H})^{2}}\,{\nabla}^{i}{\nabla}_{\left(i\right.}V_{\left.j\right)}=4\pi T\zeta_{j}-\partial_{j}p\,.
\end{align}
exactly as in \eqref{stokeslbulk}.

To summarise, at leading order in the $\epsilon$ expansion,
after applying a DC thermal gradient source
parametrised by the closed one-form $\zeta$, we can obtain the full, local stress tensor response, given in \eqref{stresspert}, after solving the Stokes equations \eqref{sttwo} on the curved manifold with metric $h_{ij}$.
Notice, in particular, from the expression for the stress tensor given in \eqref{stresspert} we find
that the local heat current of the dual field theory is given by
\begin{align}
Q^i_{QFT}&=-\sqrt{h}T^i{}_t\,, \nn
&=4\pi Tr_H^{D-4}\sqrt{h}h^{ij}V_j\,,\nn
&=4\pi Tr_H^{D-4}\epsilon^{-2}\sqrt{h}h^{ij}v^{(0)}_j\,,
\end{align}
where the index was raised using the inverse of the perturbed metric \eqref{pbmet}. 
This is just the local heat current on the horizon given in \eqref{qexp} and hence we have demonstrated that to leading order $Q^i_{QFT}(x)=Q^i_{BH}(x)$ as stated in \eqref{localcurs}.

In the previous sub-section we discussed how certain sub-leading terms in the $\epsilon$ expansion are required to 
cancel divergences at the horizon. In appendix \ref{log} we show that these lead to the following sub-leading contribution to
the heat currents:
\begin{align}\label{qsdiftext}
Q^{i}_{QFT}&=Q^{i}_{BH}+r_{H}^{D-3}\,(\epsilon\,r_{H})^{-2}\sqrt{h} \nabla_k \nabla^{[k}v^{i]}_{(0)}\,.
\end{align}
Observe that the last term on the right hand side is order $\epsilon^0$ because of the two spatial derivatives and hence is
lower order. Also
notice that the last term on the right hand side of \eqref{qsdiftext} is a total derivative (a magnetisation 
current, in fact) and hence this result is clearly consistent with the universal result of
\cite{Donos:2015gia} that the total heat current flux of the field theory is always the
same as the total heat current flux on the boundary, $\bar Q^i_{BH}= \bar Q^i_{QFT}$.

\section{Comparison with fluid-gravity approach}

In this section we would like to compare our linearised expansion of the response to a DC source with
the fluid-gravity approach discussed in \cite{Bhattacharyya:2008mz}.

We begin by recalling that our regular solution is given by
\begin{align}
ds^{2}&=r_{H}^{2}\,\rho^{2}\left(-u\,dt^{2}+h_{ij}\,dx^{i}dx^{j} \right)+\frac{d\rho^{2}}{\rho^{2}u}
-2r_{H}^{2}\rho^{2}u\,t\,\left( \zeta_{i}-(4\pi T)^{-1}\,\partial_{i}p\right)\,dx^{i}dt\nn
&-2\rho^{3-D}\left(V_{j}+V^{(2)}_j\right)\,dx^{j}dt
+2\,\rho^{2}\frac{\ln u}{4\pi T}\,\nabla_{(i}V_{j)}\,dx^{i}dx^{j}\nn
&+2\rho^{2}\,\left( V_{j}dt+t\,\nabla_{(i}V_{j)}\,dx^{i}\right)\,dx^{j}\,,
\end{align}
where we have explicitly added the one sub-leading term $V^{(2)}_j(\rho,x)$ which we have seen
is required to obtain a regular solution at the black hole horizon at leading order in the expansion. 
Recall that we have imposed $\nabla_i V^i=0$.

We next carry out the coordinate transformations
\begin{align}
&t\to v-\,R,\qquad R=-\int_\rho^\infty\frac{d\rho}{r_{H}\rho^{2}u}\,,\nn
&x^{i}\to x^{i}-r_{H}^{-2}\,v\,V^{i}\,,
\end{align}
to obtain the metric
\begin{align}
ds^{2}=&r_{H}^{2}\rho^{2}\,\left(-dv^{2}-2v\,\left( \zeta_{j}-(4\pi T)^{-1}\,\partial_{j}p\right)dx^{j} dv +h_{ij}\,dx^{i} dx^{j}\right)\nn
&-2 r_Hd\rho\,\left[ -dv+\left(r_{H}^{-2}V_{j}-v\,\left( \zeta_{j}-(4\pi T)^{-1}\,\partial_{j}p\right) \right)\,dx^{j}\right]\nn
&+\rho^{-D+3}\,r_{H}^2dv\,\left[dv-2\left(r_{H}^{-2}V_{j}-v\,\left( \zeta_{j}-(4\pi T)^{-1}\,\partial_{j}p\right) \right)\,dx^{j}\right]\nn
&+2r_H\rho^{2}F(\rho)\left(r_H^{-2}\nabla_{(i}V_{j)}\right)\,dx^{i} dx^{j}\nn
&+2r_{H}^{2}\rho^{2}u R\,\left[\left( \zeta_{j}-(4\pi T)^{-1}\,\partial_{j}p\right)-\frac{\rho^{1-D}}{r_H^2 u R}V_j^{(2)} \right]dx^{j} \left( dv- \frac{d\rho}{r_{H}\rho^{2}u}\right)\,,
\end{align}
where
\begin{align}
F(\rho)\equiv r_H\left( \frac{\ln u}{4\pi T}-R\right)\,.
\end{align}
We note that this function $F$ is the same function defined below equation (4.1) of \cite{Bhattacharyya:2008mz}.

We now perform another coordinate transformation
\begin{align}
&x^{i}\to x^{i}- h^{ij}\int_\rho^\infty d\rho\left[  \frac{R}{r_H\rho^2}\left( \zeta_{j}-(4\pi T)^{-1}\,\partial_{j}p\right) -\frac{\rho^{-1-D}}{r_H^3 u}  V_j^{(2)}    \right]  \,,    
\end{align}
to finally write our metric in the form
\begin{align}\label{eq:armet_1s}
ds^{2}=&r_{H}^{2}\rho^{2}\,\left(-dv^{2}-2v\,\left( \zeta_{j}-(4\pi T)^{-1}\,\partial_{j}p\right)dx^{j} dv +h_{ij}\,dx^{i} dx^{j}\right)\notag\\
&-2 r_H\,d\rho\,\left[ -dv+\left(r_{H}^{-2}V_{j}-v\,\left( \zeta_{j}-(4\pi T)^{-1}\,\partial_{j}p\right) \right)\,dx^{j}\right]\notag\\
&+\rho^{-D+3}\,r_{H}^2dv\,\left[dv-2\left(r_{H}^{-2}V_{j}-v\,\left( \zeta_{j}-(4\pi T)^{-1}\,\partial_{j}p\right) \right)\,dx^{j}\right]\nn
&+2r_H\rho^{2}F(\rho)\left(r_H^{-2}\nabla_{(i}V_{j)}\right)\,dx^{i} dx^{j}\nn
&+2r_{H}^{2}\rho^{2}u R\,\left[\left( \zeta_{j}-(4\pi T)^{-1}\,\partial_{j}p\right)-\frac{\rho^{1-D}}{r_H^2 u R}V_j^{(2)} \right]dx^{j} dv\,,
\end{align}
where here we have dropped covariant derivatives of $\zeta, V^{(2)}$ and second derivatives of $p$ as these would be even higher order
corrections. Notice that as we approach the horizon at $\rho=1$ there is a cancellation in the last line after
using \eqref{veetwo} as well as going on-shell by imposing the Stokes equations \eqref{stokeslbulk}, as we already discussed. 

We now consider the formalism of \cite{Bhattacharyya:2008mz}. 
The basic idea is construct an expansion for the $D$-dimensional metric using a boundary $(D-1)$-vector $u^\mu$ combined with a radial coordinate $r$. An expansion for the boundary stress tensor, $T^{\mu\nu}$, is constructed in terms of $u^\mu$ and the boundary metric, $h_{\mu\nu}$,
and then it is shown that the Ward identity $\nabla_\mu T^{\mu\nu}=0$ implies that the bulk Einstein equations are solved order
by order in the expansion. An elegant feature of the analysis in \cite{Bhattacharyya:2008mz} is the use of a Weyl covariant formalism.

Prompted by \eqref{eq:armet_1s}, let us consider the boundary metric to be given by
\begin{align}\label{eq:armet_2}
h_{\mu\nu}dx^\mu dx^\nu=-dv^{2}-2v\,\left( \zeta_{j}-(4\pi T)^{-1}\,\partial_{j}p\right)dx^{j} dv +h_{ij}\,dx^{i} dx^{j}\,,
\end{align}
and we note that we should identify $h_{\mu\nu}$ with $g_{\mu\nu}$ in the notation of \cite{Bhattacharyya:2008mz}.
With some foresight we choose the $(D-1)$-dimensional boundary fluid vector $u^\mu$ to have components
\begin{align}
u_{v}=-1\,,\qquad
u_{j}=-v\left( \zeta_{j}-(4\pi T)^{-1}\,\partial_{j}p\right)+\delta u_j\,.
\end{align}
It is now straightforward to calculate the components of the Weyl gauge-field $A_\mu$, defined in eq. (2.2) of
\cite{Bhattacharyya:2008mz}, and we find, to the order we are working to,
\begin{align}
A_{v}=\frac{1}{D-2}\nabla_i\delta u^i\,,\qquad
A_{j}=- \left( \zeta_{j}-(4\pi T)^{-1}\,\partial_{j}p\right)\,.
\end{align}
We can calculate the Weyl covariant derivative of $u^\mu$, as defined in eq. (2.3) of \cite{Bhattacharyya:2008mz}
and after writing ${\cal D}_\mu u_\nu=\sigma_{\mu\nu}+\omega_{\mu\nu}$, we obtain the 
the symmetric shear strain tensor $\sigma_{\mu\nu}$ and the antisymmetric vorticity tensor $\omega_{\mu\nu}$.
The components of $\sigma_{\mu\nu}$ are given by
\begin{align}
\sigma_{vv}=0,\qquad\sigma_{vi}=0,\qquad
\sigma_{ij}=\nabla_{(i} \delta u_{j)}-\frac{h_{ij}}{D-2}\nabla_k\delta u^k\,,
\end{align}
where the covariant derivative on $\delta u$, here and below, is with respect to $h_{ij}$.
The only non-vanishing component of the vorticity tensor is given by $\omega_{ij} = \nabla_{ [ i }\delta u_{ j ] }$, but it does not contribute to the hydrodynamics at the order we are working with. 

We can now calculate the stress tensor as given eq. (4.6) of \cite{Bhattacharyya:2008mz}. Again to the order we are working with
we have
\begin{align}
T_{\mu\nu}=p(h_{\mu\nu}+(D-1)u_\mu u_\nu)-2\eta\sigma_{\mu\nu}\,,
\end{align}
where $p=r_H^{D-1}$ and constant\footnote{The formalism of \cite{Bhattacharyya:2008mz} is Weyl covariant. In particular, we can make a Weyl transformation of the boundary metric \eqref{eq:armet_2} to introduce a non-vanishing $h_{tt}$. 
Under this Weyl transformation the stress tensor \eqref{stresscs} transforms with a Weyl factor and it may seem that the shear viscosity becomes spatially dependent. However, we will still be led to exactly the same Navier-Stokes equations with constant
shear viscosity. This is consistent with the earlier discussion of the Weyl invariance of the black hole horizon metric
given in \eqref{bhhorgen}.} shear viscosity $\eta=r_H^{D-2}$. Explicitly we obtain the components
\begin{align}\label{stresscs}
T_{vv}&=r_H^{D-1}(D-2)\,,\nn
T_{vi}&=r_H^{D-1}[  (D-2)v\left( \zeta_{j}-(4\pi T)^{-1}\,\partial_{j}p\right) - (D-1)\delta u_i]\,,\nn
T_{ij}&=r_H^{D-1}h_{ij}-2 r_H^{D-2}\left(\nabla_{(i} \delta u_{j)}-\frac{h_{ij}}{D-2}\nabla_k\delta u^k\right)\,.
\end{align}
The Ward identity $\nabla_\mu T^{\mu\nu}=0$ now gives, at leading order, both the incompressibility condition and
the linearised Navier-Stokes equations:
\begin{align}\label{sdp}
\nabla_i \delta u^i=0,\qquad
-2\,{\nabla}^{i}{\nabla}_{\left(i\right.}\delta u_{\left.j\right)}=4\pi T\zeta_{j}-\partial_{j}p\,.
\end{align}

Given this data, we can now obtain the bulk `fluid-gravity' metric given by eq. (4.1) of \cite{Bhattacharyya:2008mz}. To
the order we are considering, and after writing
\begin{align}
\delta u_i =r_H^{-2}V_i\,,
\end{align}
we find that it takes the form
\begin{align}\label{lastlll}
ds^{2}=&r_H^2\rho^2\,\left(-dv^{2}-2v\,\left( \zeta_{j}-(4\pi T)^{-1}\,\partial_{j}p\right)dx^{j} dv +h_{ij}\,dx^{i} dx^{j}\right)\notag\\
&-2r_H\,d\rho\,\left[ - dv+\left(r_H^{-2} V_{j}- v\,\left( \zeta_{j}-(4\pi T)^{-1}\,\partial_{j}p\right) \right)\,dx^{j}\right]\notag\\
&+\rho^{-D+3}r_H^{2}dv\,\left[dv-2 \left(r_H^{-2} V_{j}- v\,\left( \zeta_{j}-(4\pi T)^{-1}\,\partial_{j}p\right) \right)\,dx^{j}\right]\nn
&+2r_H\rho^{2}\,F(\rho)r_H^{-2} \,\nabla_{(i}V_{j)}\,dx^{i} dx^{j}\notag\\
&-2 r_{H}\rho\,\left( \zeta_{j}-(4\pi T)^{-1}\,\partial_{j}p\right) dx^{j} dv.
\end{align}
By comparing \eqref{lastlll} with \eqref{eq:armet_1s} we find precise agreements in the first four lines. The 
difference in the last line will be accounted for, on shell, by contributions coming from higher order terms\footnote{For example, there will be contributions coming from the last term in the third line of eq. (4.1) of \cite{Bhattacharyya:2008mz}, proportional
to $u_{(\mu}P^\lambda_{\nu)}\cal{D}_\alpha\sigma^\alpha{}_\lambda$ that are proportional to $\nabla^k\nabla_{(k}V_{i)} dv dx^i$
and hence proportional to $\left( \zeta_{j}-(4\pi T)^{-1}\,\partial_{j}p\right) dx^{j} dv$ on-shell.}
in the expansion.

To close this section we present an equivalent way of describing the DC linear response from the fluid-gravity perspective.
Consider, carrying out the simple coordinate transformation on the boundary metric \eqref{eq:armet_2}, $v\to (1-\psi)v$ with
$\psi$ a locally defined function satisfying 
\begin{align}
d\psi=\left( \zeta_{i}-(4\pi T)^{-1}\,\partial_{i}p\right) \,dx^{i}\,.
\end{align}
In the new coordinates the boundary metric takes the form
\begin{align}\label{eq:armet_3}
h_{\mu\nu}dx^\mu dx^\nu=-(1-2\psi)dv^{2} +h_{ij}\,dx^{i} dx^{j}\,,
\end{align}
which is another way to introduce the DC thermal gradient. The fluid velocity is then given by the more intuitive expressions
\begin{align}\label{intexs}
u_{v}=-(1-\psi)\,,\qquad
u_{j}=\delta u_{j}\,,
\end{align}
while the stress tensor takes the form
\begin{align}\label{intstress}
T_{vv}&=r_H^{D-1}(D-2)(1-2\psi),\nn
T_{vi}&=-r_H^{D-1}(D-1)\delta u_i\,,\nn
T_{ij}&=r_H^{D-1}h_{ij}-2 r_H^{D-2}\left(\nabla_{\left(i\right.} \delta u_{\left. j\right)}-\frac{h_{ij}}{D-2} \nabla_{i}\delta u^{i}\right)\,.
\end{align}
Imposing the Ward identity $\nabla_\mu T^{\mu\nu}=0$ again gives the incompressibility condition and
the linearised Navier-Stokes equations \eqref{sdp}. This fluid can then be used to construct the bulk metric, in the hydrodynamic limit, 
using the 
formulae in \cite{Bhattacharyya:2008mz}. It should be noted, however, that in these coordinate the bulk metric is constructed from 
the local function $\psi$ and the regularity of the solution is not immediately transparent. 

It is interesting to point out that if we take the boundary metric \eqref{eq:armet_3}, but allow $\psi$ to also depend on time, we can still take the fluid velocity as in \eqref{intexs} and the stress tensor is still given as in
\eqref{intstress}. The Ward identities still imply the incompressibility condition $\nabla_{i}\delta u^{i}=0$ but now we
obtain the linearised Navier-Stokes equation
\begin{align}
4\pi T\,\partial_{t}\delta u_{i}-2 \nabla^{j}\nabla_{\left(j\right.} \delta u_{\left. i\right)}+\partial_{i}p=4\pi T \zeta_{i}\,.
\end{align}
By employing the scaling discussed at the end of section \ref{crtresults}, we deduce that this equation can be used to consider time-dependence that is associated with frequencies $\omega\sim \epsilon k\sim\epsilon^2 T$. This indicates that the poles of the current-current correlator will be
order $\epsilon^2$ for fixed $T$, on the negative imaginary axis in the complex $\omega$ plane
and hence, combining with \eqref{hightscale}
we infer that as $\epsilon\to 0$ we have weak momentum dissipation and an associated Drude peak.

\section{Discussion}

In this paper we have discussed the DC response of holographic lattices in theories of pure gravity, in a hydrodynamic limit.
We have shown that by solving the linearised, covariantised Navier-Stokes equations for an incompressible fluid one can extract out the local heat currents of the dual field theory as well as determining the leading order correction. In addition we also determined the full local stress tensor response at leading order.
For simplicity we only considered static holographic lattice black hole geometries.
However, our results can be extended to the stationary case, corresponding to having local momentum current deformations in the dual CFT, using
the results of \cite{Donos:2015bxe}. In particular, by solving the generalised Navier-Stokes equations of \cite{Donos:2015bxe} with specific 
Coriolis terms, one can extract the local transport heat currents. The effects of the Coriolis term could give a sharp diagnostic to determine the presence of such magnetisation currents in real systems.

We have focussed on holographic lattices that are periodic in non-compact spatial dimensions. In particular,
in the limit that $\epsilon\to 0$ we obtain the AdS-Schwarzschild black brane geometry. Our analysis
can easily be adapted to lattices that are associated with the AdS-Schwarzschild black holes with hyperbolic
horizons. An interesting feature is that by taking suitable quotients of the hyperbolic space one can study DC thermal
conductivities on higher genus Riemann surfaces. By contrast one cannot use  
AdS-Schwarzschild black holes with spherical horizons in the same way because of the absence of one-cycles 
to set up a DC source (recall that the DC source was parametrised by a closed one-form $\zeta$).

For holographic lattices we have emphasised that the hydrodynamic limit is, in general, distinct from the 
perturbative lattices analysed in
\cite{Donos:2015gia,Banks:2015wha}. The hydrodynamic limit in this paper 
corresponds to an expansion in $k/T$ (or $k/s^{1/(D-2)}$), where $k$
is the largest wave number of the UV deformation,
while for the perturbative lattice one expands in a 
dimensionless parameter $\lambda$ associated with the amplitude of the UV deformation.
Despite the differences in the limits, there are some similarities. In this paper, by demanding regularity
of the perturbation at the black hole horizon we saw that at leading order $V_i=\epsilon^{-2}v^{(0)}_i$ and $p=\epsilon^{-1}p^{(0)}$, while
in \cite{Donos:2015gia,Banks:2015wha} there was an analogous expansion with $\epsilon$ replaced with $\lambda$.
In both cases the DC conductivity is parametrically large and there is weak momentum dissipation.
However, in the case of perturbative lattices the leading order solution to the Stokes equations is spatially
homogeneous (i.e. constant) on the torus, while in the hydrodynamic limit, the leading order solution generically has a non-trivial local structure.

We have also discussed how our results are related to the fluid-gravity approach and in particular the general results of 
\cite{Bhattacharyya:2008mz} who examined CFTs on boundary manifolds with arbitrary metrics.
Interestingly we showed that our perturbative hydrodynamic expansion differs from that of \cite{Bhattacharyya:2008mz}, off-shell,
but it is consistent on-shell, when one imposes the Ward identities. This difference arises from the way in which the expansion is organised
using either Schwarzschild coordinates or Eddington-Finklestein coordinates.

We saw that using the fluid-gravity formalism one can rather easily obtain the result that the local heat current of the dual field theory
can be obtained by solving Stokes equations. However, it should be pointed out that this approach obscures the fact that the boundary heat current is equal to the heat current at the horizon, as we have shown to be the case in section \ref{review}. 
Moreover, we note that extracting the universal result for holographic lattices
of \cite{Donos:2015gia,Banks:2015wha,Donos:2015bxe}
concerning Navier-Stokes equations on the black hole horizon, as an exact statement in holography, is highly non-trivial in the fluid-gravity expansion. In essence this is because the Ward identities are imposed at the AdS boundary in the fluid-gravity approach, while to get the Navier-Stokes equations the constraints should be imposed on the black hole horizon.
In effect, to obtain the result of \cite{Donos:2015gia,Banks:2015wha,Donos:2015bxe}, one would need to sum 
up an infinite expansion from the fluid-gravity point of view.

We also note that there is not an existing general fluid-gravity formalism that can be deployed for studying
the hydrodynamic limit of holographic lattices in more general theories of gravity coupled to various matter fields.
By contrast it is rather clear how some results of this paper can be generalised. For example, suppose we have
a theory of gravity coupled to a massless scalar field with no potential in the bulk. Such a scalar is dual to an exactly  
marginal operator in the dual CFT. If we consider holographic lattices with UV metric deformations, $h_{ij}(x)$, as well as UV scalar deformations, $\phi(x)$, 
then the metric on the horizon will be $r_H^2 h_{ij}(x)$, as we discussed in this paper, while the scalar field on the horizon
will be given by the UV function $\phi(x)$. We can now obtain the local heat currents on the horizon, and hence for the dual field theory
in the hydrodynamic limit, by solving the generalised Navier-Stokes equations on the horizon with the scalar viscous terms 
derived in \cite{Banks:2015wha}:
\begin{align}\label{stokesgeneral2}
-{2}\,{\nabla}^{i}{\nabla}_{\left(i\right.}v_{\left.j\right)}+(v^i\nabla_i\phi)\nabla_j\phi=4\pi T\zeta_{j}-\partial_{j}p\,,\qquad \nabla_i v^i=0\,,
\end{align}
and metric $r_H^2 h_{ij}$. 

Consider now a theory of gravity coupled to a scalar field that is dual to a relevant operator with dimension $\Delta<D-1$. At the AdS boundary the scalar field will behave as 
$\phi\sim \phi_s\bar\phi(x)r^{\Delta-D+1}$
where $\phi_s$ is a dimensionful source amplitude and $\bar\phi(x)$ is a dimensionless function.
By taking the hydrodynamic limit we again find a horizon metric $r_H^2h_{ij}$. 
By analysing the radial behaviour of the
scalar (as in section 3.3 of \cite{Donos:2014cya}), we can determine that at the black hole horizon the scalar field will be 
given by $c\phi_s T^{\Delta-D+1}\bar\phi(x)$, where $c$ is a numerical constant. We can then study the
DC response using \eqref{stokesgeneral2}. 

It is interesting to point out that if we consider the 
high temperature limit with both $\epsilon<<1$ and 
$\lambda\equiv \phi_s T^{\Delta-D+1}\bar\phi(x)<<1$, then providing that
there is also momentum dissipation arising from a spatially dependent metric, it is clear from
\eqref{stokesgeneral2} that the scalar field will not play a role at leading order. 
However, for holographic lattices in which the UV boundary metric deformations are trivial, 
$h_{ij}=\delta_{ij}$, but the relevant scalar field deformations are
non-trivial, the origin of the momentum dissipation comes purely from the scalar field and hence to obtain the leading order heat currents one will need to solve the Navier-Stokes equations with the scalar viscous terms. In the high temperature limit the scalar viscous terms are small, so we are now in the domain of perturbative lattices associated with weak momentum dissipation, and we can obtain the leading DC conductivity in terms of the scalar field on the horizon using the results of \cite{Donos:2015gia,Banks:2015wha}.

Finally, we can also consider the addition of gauge fields. In this case we can take the hydrodynamic limit
of holographic lattices at finite charge density, by just demanding $k/s^{1/(D-2)}<<1$ while holding fixed $\mu/s^{1/(D-2)}$.
One should then study the charged Navier-Stokes equations of \cite{Donos:2015gia,Banks:2015wha}.
This will be discussed in detail elsewhere.

\section*{Acknowledgements}

We would like to thank Simon Gentle, Michal Heller and Mukund Rangamani 
for helpful discussions. 
The work of JPG, TG and LM is supported by the European Research Council under the European Union's Seventh Framework Programme (FP7/2007-2013), ERC Grant agreement ADG 339140. The work of JPG is also supported
by STFC grant ST/L00044X/1 and EPSRC grant EP/K034456/1. JPG is also supported as a KIAS Scholar and
as a Visiting Fellow at the Perimeter Institute. The work of EB supported by an Imperial College Schr\"odinger Scholarship.

\appendix

\section{Sub-leading corrections of the linearised perturbation}\label{log}

We consider a metric of the form
\begin{align}\label{metapp}
ds^{2}=-U\,G\,\left(dt+\delta\chi\right)^{2}+\frac{F}{U}\,dr^{2}+g_{ij}dx^i dx^j\,,
\end{align}
as in \eqref{ansatz}, but now with an additional linear perturbation $\delta \chi(r,x)$.
We would like to understand the behaviour of $\delta\chi$ as a perturbative expansion about the high temperature background
solution \eqref{eq:gen_back22} as well as the sub-leading perturbative corrections that are polynomial in $\epsilon$. 
Later, when we combine this with the DC thermal gradient source as in \eqref{eq:pert_l1}, we will then see an elegant interplay between solutions of the Stokes equations and regularity of the combined perturbation
at the horizon.

Without loss of generality we work in a coordinate system where $\delta\chi_r=0$ (this can be achieved via the coordinate transformation $t\to t+f(r,x)$) and take $\delta\chi=\delta\chi_i(r,x) dx^i$. The equation of motion of $\delta\chi$ is then given by
\begin{align}\label{eq:chi_eqn}
\partial_r(U^{2}G^{3/2}F^{-1/2}\sqrt{g_d}\,g^{ij}\,\partial_r\delta\chi_{j} )+\partial_{k}[U\,G^{3/2}F^{1/2}\sqrt{g_d}\,g^{kl}g^{ij}\,(\partial_{l}\delta\chi_{j}-\partial_{j}\delta\chi_{l} )]=0\,,
\end{align}
as well as
\begin{align}\label{eqdiv}
\partial_{i}(U^{2}\,G^{3/2}F^{-1/2}\sqrt{g_d}\,g^{ij}\,\partial_r\delta\chi_{j} )=0\,,
\end{align}
where $\sqrt{g_d}$ is the volume element associated with $g_{ij}$.
One solution of these equations is to take an arbitrary closed form given by  $\delta\chi =\chi_{i}(x) dx^i$. This solution gives rise to a source in the dual field theory and is not what we are interested in here. 

We now consider the background solution as a derivative expansion in the high-temperature limit:
\begin{align}\label{eq:bac_lambda_exp}
U&=r_{H}^2\,\rho^{2}\,u(\rho)=r_{H}^2\,\rho^{2}\,(1-\rho^{1-D})\,,\nn
G&=\left(1+\epsilon^{2}\,G^{(2)}(\rho,x) +\cdots\right)\,,\nn
F&=\left( 1+\epsilon^{2}\,F^{(2)}(\rho,x)+\cdots \right)\,,\nn
g_{ij}&=r^{2}_{H}\,\rho^{2}\,\left( h_{ij}(x)+\epsilon^{2}\,h_{ij}^{(2)}(\rho,x)+\cdots\right)\,,
\end{align}
where $\epsilon=k/T$, with $k$ the largest wavenumber of the background, 
and $r=r_H\rho$. We note that, in general, there will also be corrections that
are non-perturbative in $\epsilon$ which we will ignore.
As before $ h_{ij}(x)$ is the UV deformation of the metric and since we want this to be the only deformation
of the dual field theory, we demand that $G^{(2)}$, $F^{(2)}$ and $h_{ij}^{(2)}$ all vanish as $\rho\to\infty$.
Note that regularity at the horizon imposes $F^{(2)}|_{\rho=1}=G^{(2)}|_{\rho=1}$ and that the metric on the horizon is
$r^{2}_{H}( h^{(0)}_{ij}+\epsilon^{2}\,h_{ij}^{(2)}|_{\rho=1}+\cdots)$.

We want to solve \eqref{eq:chi_eqn}, \eqref{eqdiv} perturbatively in $\epsilon$ by postulating an expansion of the form
\begin{align}\label{eq:chi_exp}
\delta\chi_{i}=\epsilon^{\nu}\,( \chi_{i}^{(0)}+\epsilon^{2}\,\chi_{i}^{(2)}+\cdots)\,,
\end{align}
with $\nu$ an exponent that we will eventually have to fix. It is clear that the homogeneous, linear equation \eqref{eq:chi_eqn} cannot fix this exponent. It must be fixed by an inhomogeneous constraint involving the
boundary and or the horizon. As we will see it is fixed by ensuring the perturbation is regular at the horizon when it is combined
with the perturbation associated with the DC thermal gradient source as in \eqref{eq:pert_l1}.

To carry out the expansion in $\epsilon$ we note that any spatial derivative $\partial_i$ is of order
$\epsilon$. Now the second term in \eqref{eq:chi_eqn} comes with two spatial derivatives so it
is order $\epsilon^2$. Thus, at leading order in $\epsilon$
we have the simple ODE
\begin{align}
\partial_{\rho}(\rho^{D}\,u^{2}\,\partial_{\rho}\chi_{i}^{(0)})=0\,,
\end{align}
and solutions are given by
\begin{align}\label{zorderone}
\chi_{i}^{(0)}(\rho,x)=c^{(0)}_{i}(x)+\frac{\rho^{1-D}}{r^{2}_{H}\,u(\rho)}\,v_{i}^{(0)}(x)\,.
\end{align}
To ensure that there are no additional sources at
infinity we set
\begin{align}\label{zordertwo}
c^{(0)}_{i}=0\,.
\end{align}
From equation \eqref{eqdiv} we also have
\begin{align}\label{zorderthree}
\nabla_i v^{i}_{(0)}=0\,,
\end{align}
where $\nabla$ is the Levi-Civita connection for the spatial metric $h_{ij}$ and indices have been raised with $h^{ij}$.

At second order in $\epsilon$, equation \eqref{eq:chi_eqn} implies
\begin{align}\label{eq:eq:chi_eqn_sec_order}
&r_{H}^{D-2}\,\sqrt{h}\,h^{ij}\partial_{\rho}(\rho^{D}u^{2}\,\partial_{\rho}\chi^{(2)}_{j})+r_{H}^{D-2}\,\partial_{\rho}(\rho^{D}u^{2}\,\sqrt{h}\,N^{ij}\,\partial_{\rho}\chi^{(0)}_{j})\notag\\
&\qquad\qquad\qquad +\epsilon^{-2}\,r_{H}^{D-4}\,\rho^{D-4}\,\partial_{k}(u\,\sqrt{h}\,h^{kl}\,h^{ij}\,(\partial_{l}\chi_{j}^{(0)}-\partial_{j}\chi_{l}^{(0)}))=0\,,
\end{align}
where we defined the matrix $N^{ij}(\rho,x)$:
\begin{align}\label{endef}
N^{ij}=h^{ij}\,\left[\frac{3}{2}\,G^{(2)}-\frac{1}{2}\,F^{(2)}\right]+\frac{1}{2}\,h^{ij}\,h^{kl}h_{(2)}{}_{kl}(\rho,x^{i})-h_{(2)}^{ij}\,,
\end{align}
and the indices on the last term have been raised with $h^{ij}$.
After substituting the zeroth order solution \eqref{zorderone}, \eqref{zordertwo} we can rewrite \eqref{eq:eq:chi_eqn_sec_order} as
\begin{align}\label{eq:eq:chi_eqn_sec_ordertwo}
&h^{ij}\partial_{\rho}(\rho^{D}u^{2}\,\partial_{\rho}\chi^{(2)}_{j})
-\frac{D-1}{r_{H}^{2}}\,\partial_{\rho}N^{ij}\,v^{(0)}_{j}
+\frac{2\rho^{-3}}{r_H^2(\epsilon r_H)^2}\, \nabla_k \nabla^{[k}v^{i]}_{(0)}=0\,.
\end{align}
The general solution to this equation is of the form
\begin{align}\label{eq:sec_order_sol}
\chi_{i}^{(2)}(\rho,x)=c^{(2)}_{i}(x)+\frac{\rho^{1-D}}{r^{2}_{H}\,u(\rho)}\,v_{i}^{(2)}(x)+q_{i}^{(2)}(\rho,x)\,.
\end{align}
The first two terms are the solutions to the homogeneous equation (with $v^i_{(0)}=0$) and the third term is a particular solution of the inhomogeneous equation. The solution that we are interested in, to ultimately ensure that we have non-singular behaviour near the horizon, will be such that $q_{j}^{(2)}$ has no $(\rho-1)^{-1}$ term close to the horizon, but instead a $\log(\rho-1)$ behaviour.
The function $c^{(2)}_{j}$ is fixed so that we don't have a source at infinity at the given order in the $\epsilon$ expansion; 
as we will see this implies that $c^{(2)}_{j}=0$. We will see that these requirements uniquely fix $q_{j}^{(2)}$. On the other hand, the new function of integration $v^{(2)}_i(x)$ will be fixed at the next order in the expansion, a point we will return to later.

To proceed, we demand that in an expansion close to horizon the leading term of $q_{j}^{(2)}$ is given by
\begin{align}\label{eq:q_hlog}
q_{j}^{(2)}(\rho,x^{i})=q_{j}(x^{i})\,\log(\rho-1)+\cdots\,.
\end{align}
From equation \eqref{eq:eq:chi_eqn_sec_ordertwo} we deduce
\begin{align}
(4\pi T)^{2}\,h^{ij}\,q_{j}=(D-1)\,v^{(0)}_{j}\,\partial_{\rho}N^{ij}|{}_{\rho=1}
-2(\epsilon\,r_{H})^{-2}\, \nabla_k \nabla^{[k}v^{i]}_{(0)}\,.
\end{align}
Now a key point is that the leading term in the expansion of the matrix $N^{ij}$ near the horizon is actually fixed 
by horizon data. This result can be extracted\footnote{To compare we have $r^{there}=r_H(\rho-1)$.
Then in the notation of eq. (2.5) of \cite{Banks:2015wha} we have,
with objects on the left hand side in the notation of \cite{Banks:2015wha},
 $U^{(1)}=4\pi T (4-D)/(2r_H)$, $G^{(0)}=F^{(0)}=1+\dots$, $G^{(1)}=(r_H)^{-1}\epsilon^2\partial_\rho G^{(2)}|_{\rho=1}+\dots$ 
 and
$F^{(1)}=(r_H)^{-1}\epsilon^2\partial_\rho F^{(2)}|_{\rho=1}+\dots$. We then use (D.2) and (D.4) of \cite{Banks:2015wha} to
get \eqref{ennresult}.} from calculations presented in \cite{Banks:2015wha} and we have
\begin{align}\label{ennresult}
\partial_{\rho}N^{ij}|_{\rho=1}=-\frac{2}{(D-1)}\,\frac{1}{(\epsilon\,r_{H})^{2}}\,R^{ij}\,,
\end{align}
where $R_{ij}$ is the Ricci tensor of the $d$-dimensional UV metric $h_{ij}$, and again the indices have been raised
using $h^{ij}$. Using \eqref{zorderthree} we can then deduce
\begin{align}
q_{j}=-\frac{2}{(4\pi T)^{2}}\,\frac{1}{(\epsilon\,r_{H})^{2}}\,{\nabla}^{i}{\nabla}_{\left(i\right.}v^{(0)}{}_{\left.j\right)}\,.
\end{align}

Having established how the expansion of the perturbation $\delta\chi$ works, at this point we now recall that the full perturbation that we are interested in also has the time dependent piece
given in \eqref{eq:pert_l1}. By switching to the ingoing coordinate,
$v=r_H(t+\ln u/(4\pi T))$, we see that the $\log(\rho-1)$ factors will
be eliminated in the full perturbation provided that we choose the exponent in \eqref{eq:chi_exp} to be
\begin{align}
\nu=-2\,,
\end{align}
and, in addition, demand that $v^{(0)}_i$ satisfies the linearised Navier-Stokes equation:
\begin{align}\label{stokesl}
-\frac{2}{(\epsilon r_{H})^{2}}\,{\nabla}^{i}{\nabla}_{\left(i\right.}v^{(0)}_{\left.j\right)}=4\pi T\zeta_{j}-\partial_{j}p\,,
\end{align}

Now a consideration of \eqref{metapp}, \eqref{eq:bac_lambda_exp}, \eqref{eq:chi_exp}
and comparing with \eqref{eq:pert_l3}, reveals that the combined perturbation given in \eqref{eq:pert_l1}-\eqref{eq:pert_l4}
is regular at the horizon, to leading order, provided that we have
\begin{align}
V_{i}(x)=\epsilon^{-2}\,v_{i}^{(0)}(x)\,,
\end{align}
with $v_{i}^{(0)}$ satisfying \eqref{stokesl}. 
Furthermore, since we know that, in general, the Stokes equations are satisfied at the horizon we learn that we must have
\begin{align}
p=\epsilon^{-1}p^{(0)}+\dots\,.
\end{align}
and that $4\pi T\zeta$ is the same order as $p^{(0)}$ and $v_{i}^{(0)}$.

Having established the behaviour at the horizon, we now give the full integrated expression for $q_{j}^{(2)}(\rho,x)$:
\begin{align}\label{eqn:q_exp}
q_{j}^{(2)}=&\frac{D-1}{r_{H}^{2}}\,h{}_{ij}\,v_{l}^{(0)}\,\int_{\infty}^{\rho}\frac{N^{il}(\rho^{\prime},x)}{u^{2}(\rho^{\prime})}\,\rho^{\prime}{}^{-D}\,d\rho^{\prime}+
\frac{\rho^{1-D}}{r_H^2u(\rho)}h{}_{ij}\,v_{l}^{(0)}\,N^{il}|_{\rho=1}\notag\\
&+\left( \epsilon\,r_{H}\right)^{-2}\left[\,\left(\int_{\infty}^{\rho} \frac{\rho^{\prime}{}^{-D-2}}{u^{2}(\rho^{\prime})}d\rho^{\prime}\right)\,
+\frac{1}{(D-1)}\,\frac{\rho^{1-D}}{u(\rho)}\,\right] \nabla^k \nabla_{[k}v_{j]}^{(0)}\,.
\end{align}
Note that the radial dependence in $q_{j}^{(2)}$ can be made explicit provided that we can find the back-reaction of the background at order $\epsilon^2$ that is packaged in $N^{ij}$. Observe that the last term in the first line and the last line of  \eqref{eqn:q_exp} are simply solutions of the homogeneous equation (their radial dependence is the same with the second term of \eqref{eq:sec_order_sol}) 
with the corresponding function of integration chosen such that we get the behaviour \eqref{eq:q_hlog} without a potential $(\rho-1)^{-1}$ term, as we mentioned earlier. Furthermore one can check that 
the leading behaviour close to the horizon of the expression above is indeed given by \eqref{eq:q_hlog}.

From this expression we can now extract the asymptotic behaviour at $\rho=\infty$. We first note that $N^{ij}$ goes to zero close to the boundary since it is constructed from higher order corrections to the background and by assumption these
do not change the conformal boundary metric.
We find 
\begin{align}\label{eq:q_asympt}
\chi_{j}^{(2)}=&\rho^{1-D}\,r_{H}^{-2}\,v^{(2)}_{j}+\rho^{1-D}r_H^{-2}\,h{}_{ij}\,v_{l}^{(0)}\,N^{il}|_{\rho=1}
+\frac{\rho^{1-D}}{(D-1)}\,\left( \epsilon\,r_{H}\right)^{-2}\, \nabla^k \nabla_{[k}v_{j]}^{(0)}+\cdots\,,
\end{align}
where we have set $c^{(2)}_{i}=0$, which we now see does indeed correspond to having vanishing source term for the perturbation.
 
At this point one might wonder how $v^{(2)}_{j}$ is fixed. The next order in the expansion will include a function $q^{(4)}_j(\rho,x)$. Once
again, near the horizon the $(\rho-1)$ behaviour will be eliminated leaving $\log(\rho-1)$ behaviour. Regularity at the horizon will imply that this is in turn be fixed by the corrected Navier-Stokes equation at the horizon (note that the perturbation \eqref{eq:pert_l1} is also corrected because $g_{tt}$ will also receive corrections). 
While this procedure can be carried out in detail, for the main results we want to present here, we will not need to.
An additional point is that that we also need to satisfy \eqref{eqdiv} at next order. This condition reads
\begin{align}
\partial_i\left(\sqrt{h}h^{ij}\partial_\rho\chi^{(2)}_j+\sqrt{h}N^{ij}\partial_\rho\chi^{(0)}_j\right)=0\,.
\end{align}
After substituting the expressions in and using the fact that
$\nabla_i\nabla_j \nabla^{[i}v^{j]}_{(0)}=0$, we find that we must impose
\begin{align}
\partial_i\left(\sqrt{h}h^{ij}v^{(2)}_j+\sqrt{h}N^{ij}|_{\rho=1}v^{(0)}_j\right)=0\,.
\end{align}

At this point we have established that this perturbation has an expansion which can be written
\begin{align}\label{fexac}
\delta g_{ti}(\rho,x)&=-UG\delta\chi_i\nn
&=-\epsilon^{-2}\rho^{3-D}\left(v^{(0)}_i(x)+\epsilon^2V^{(2)}_i(\rho,x) +\dots\right)\,,
\end{align}
where
\begin{align}\label{fexac2}
V^{(2)}_i(\rho,x) =G^{(2)} v^{(0)}_i(x)+v^{(2)}_i(x) + \frac{r_H^2 u}{\rho^{1-D}}q^{(2)}_i(\rho,x)  \,,
\end{align}
with $q^{(2)}_i(\rho,x)$ given in \eqref{eqn:q_exp} and an expression for 
$v^{(2)}_i(x)$ can be explicitly obtained by continuing to higher orders. We have also shown that as we approach the horizon
we have
\begin{align}
V^{(2)}_i(\rho,x) \to \frac{r_H^2}{(4\pi T)^2}u \log(\rho-1)\left( -\frac{2}{(\epsilon r_{H})^{2}}\,{\nabla}^{i}{\nabla}_{\left(i\right.}v^{(0)}_{\left.j\right)}   \right), \qquad \rho\to 1\,.
\end{align}
It is also worth noting that $v_i(x)$ that appears in the general Stokes equations on the black hole horizon (see \eqref{stokesgeneral}) is given by $v_i(x)= -\delta g_{ti}|_{\rho=1}$.

An important objective is to obtain the local heat current density of the dual field  theory $Q^{i}_{QFT}$. We conclude this appendix  by showing how that at leading order in the
expansion we have $Q^{i}_{QFT}(x)=Q^{i}_{BH}(x)$, as well as indicating the structure
of the sub-leading corrections.
We first return to \eqref{eq:chi_eqn} and observe that this equation can be rewritten as
\begin{align}
\partial_\rho Q^i=r_H\partial_{k}[ U\,G^{3/2}F^{1/2}\sqrt{g_d}\,g^{kl}g^{ij}\,(\partial_{l}\delta\chi_{j}-\partial_{j}\delta\chi_{l} )]\,,
\end{align}
where $Q^i=-(r_H)^{-1}U^{2}G^{3/2}F^{-1/2}\sqrt{g_d}\,g^{ij}\,\partial_{\rho}\delta\chi_{j}$ is the bulk thermal current, defined in (3.20) of
\cite{Banks:2015wha}. By integrating in the radial direction we deduce that
\begin{align}
Q^{i}_{QFT}-Q^{i}_{BH}=r_H\int_{\rho=1}^{\infty}d\rho\, \partial_{k}[U\,G^{3/2}F^{1/2}\sqrt{g_d}\,g^{kl}g^{ij}\,(\partial_{l}\delta\chi_{j}-\partial_{j}\delta\chi_{l} )]\,,
\end{align}
which for the current densities gives, up to second order in $\epsilon$,
\begin{align}\label{qsdif}
Q^{i}_{QFT}-Q^{i}_{BH}&=r_{H}^{D-3}\,\epsilon^{-2}\,\int_{\rho=1}^{\infty}d\rho\, u\,\rho^{D-4}\,\partial_{k}( \sqrt{h}\,h{}^{kl} h{}^{ij}\,(\partial_{l}\chi^{(0)}_{j}-\partial_{j}\chi^{(0)}_{l} ))\,,\nn
&=2r_{H}^{D-3}\,(\epsilon\,r_{H})^{-2}\sqrt{h}\,\int_{\rho=1}^{\infty}d\rho\, \,\rho^{-3}\, \nabla_k \nabla^{[k}v^{i]}_{(0)}\,,\nn
&=r_{H}^{D-3}\,(\epsilon\,r_{H})^{-2}\sqrt{h}\, \nabla_k \nabla^{[k}v^{i]}_{(0)}\,.
\end{align}
Observe that because of the two spatial derivatives the term on the right hand side of \eqref{qsdif}  is of order $\epsilon^0$.
Thus, we conclude that at leading order the heat current at the horizon is the same as at that of the dual field theory 
and moreover we also
have obtained the leading order correction in the $\epsilon$ expansion.
We also note that the heat current at the horizon is given by
\begin{align}
Q^i_{BH}= r_H^{d-3}4\pi T\sqrt{h}\left[\epsilon^{-2} h^{ij}v^{(0)}_j
+h^{ij}v^{(2)}_j+N^{ij}|_{\rho=1}v^{(0)}_j\right]\,,
\end{align}
where the sub-leading corrections involve corrections to the background
via $N^{ij}|_{\rho=1}$ (see \eqref{endef} and \eqref{eq:bac_lambda_exp}) as well
as the sub-leading terms in the perturbation, $v^{(2)}_j$, which can be obtained by
the method discussed above.

Finally, we note 
that since the right hand side of \eqref{qsdif} is a total derivative, this result is clearly consistent with the universal result of
\cite{Donos:2015gia} that the total heat current flux of the field theory is always the
same as the total heat current flux on the boundary, $\bar Q^i_{BH}= \bar Q^i_{QFT}$.


\providecommand{\href}[2]{#2}\begingroup\raggedright\endgroup

\end{document}